\documentclass[usenatbib,onecolumn]{mn2e}
\def\gtrsim{\ \raise 3pt \hbox{$>$} \kern -6.5pt \raise -2pt \hbox{$\sim$}\ }
\def\lesssim{\ \raise 3pt \hbox{$<$} \kern -6.5pt \raise -2pt \hbox{$\sim$}\ }
\usepackage{graphicx,natbib,color,amsmath,subfigure}
\newcommand{\vp}{\mbox{\boldmath $p$}}         
\newcommand{\vk}{\mbox{\boldmath $k$}}         
\newcommand{\vv}{\mbox{\boldmath $v$}}         
\newcommand{\vu}{\mbox{\boldmath $u$}}         

\newcommand{\vE}{\mbox{\boldmath $E$}}         
\newcommand{\vr}{\mbox{\boldmath $r$}}         
\newcommand{\vj}{\mbox{\boldmath $j$}}         
\newcommand{\vJ}{\mbox{\boldmath $J$}}         
\newcommand{\vB}{\mbox{\boldmath $B$}}         
\newcommand{\vD}{\mbox{\boldmath $D$}}         
\newcommand{\vb}{\mbox{\boldmath $b$}}         
\newcommand{\vf}{\mbox{\boldmath $f$}}         
\newcommand{\const}{{\rm const}}

\def\cE{{\cal E}}

\def\q{\qquad}

\newcommand{\vcF}{\mbox{\boldmath{$\cal F$}}}
\newcommand{\vcO}{\mbox{\boldmath{$\cal O$}}}

\newcommand{\apj}    {{ApJ}}
\newcommand{\apjl}   {{ApJ}}

\newcommand{\aap}    {{A\&A}}

\newcommand{\jgr}   {{Journ. Geo. Res.}}

\newcommand{\solphys} {{Solar Phys.}}
\newcommand{\ssr}    {{Space Science Rev.}}

\newcommand{\pa}{\mbox{$\partial$}}

\begin{document}
\title{Stochastic Particle Acceleration by Helical Turbulence in Solar Flares}
\author[Gregory D. Fleishman and Igor N. Toptygin]{Gregory D. Fleishman$^{1,2,3}$ and Igor N. Toptygin$^{4}$\\
$^{1}${Center For Solar-Terrestrial Research, New Jersey Institute of Technology, Newark, NJ 07102}\\
$^{2}${Ioffe Physico-Technical Institute, St. Petersburg 194021, Russia}\\
$^{3}${Central Astronomical Observatory at Pulkovo of RAS, Saint-Petersburg
196140, Russia}\\
$^{4}${State Polytechnical University, St.Petersburg, 195251, Russia}}

\date{Accepted 2012 July ... Received 2012 July ...; in original form 2012 July 27}

\pagerange{\pageref{firstpage}--\pageref{lastpage}} \pubyear{2012}

\maketitle

\label{firstpage}

\begin{abstract}
Flaring release of magnetic energy in solar corona is only possible if the magnetic field deviates from a potential one.
We show that the linear MHD modes excited on top of the non-potential magnetic field possess a nonzero kinetic helicity. Accordingly,
this 
necessarily results in a noticeable kinetic helicity of the turbulence, composed of these linear modes with various scales and random phases,  generated at the flare site by the primary energy release, which may be important for many applications. In particular, a nonzero turbulence helicity has a potentially strong effect on the particle acceleration because the helical component of the turbulence induces 
a mean regular large-scale (DC) electric field capable of directly accelerating the charged particles in addition to the commonly considered stochastic turbulent electric field.
In this paper, we derive the kinetic helicity density of the linear MHD modes excited on top of a twisted large-scale magnetic field, estimate the corresponding turbulence helicity and take its effect on stochastic particle acceleration by the turbulence into consideration; in particular, we compare this induced mean electric field with the electron and estimated effective ion Dreicer fields.
We have discovered that this, so far missing but highly important, ingredient of the turbulence at the flare site can be responsible for the thermal-to-nonthermal energy partition in flares by controlling the process of particle extraction from the thermal pool and formation of the seed particle population to be then stochastically accelerated to higher energies. In addition, it   is naturally consistent with such puzzling flare manifestations as spatial separation of electron and proton emission sites, electron beam formation, and enrichment of the accelerated particle population by $^3He$ and other rare ions.
\end{abstract}

\begin{keywords}
Sun: flares---acceleration of
particles---turbulence---diffusion---magnetic fields---Sun:
radio radiation
\end{keywords}

\section{Introduction}

Stochastic acceleration by turbulence is currently believed to play a dominant role in charged particle acceleration in impulsive solar flares. Indeed, it has been recognized long ago \citep{Petrosian92, Petrosian94, Miller_etal_1997, Miller_1998} that {models of} stochastic acceleration {of charged particles by a postulated} MHD or whistler wave ensemble {are overall} consistent with global manifestations of the acceleration process such as total energy content of the accelerated particles, the spectrum shape, the highest energy of the accelerated particles, and the acceleration time. Furthermore, the stochastic acceleration was shown to be   consistent with a number of  more detailed acceleration properties \citep{Grigis_Benz_2006, Byk_Fl_2009}, e.g., soft-hard-soft (SHS) spectrum evolution established in detail based on the RHESSI observations \citep{Grigis_Benz_2004}. In addition, radio detection of the very acceleration region in a flare \citep{Fl_etal_2011} is suggestive in favor of a stochastic acceleration process with roughly energy independent electron residence time in the acceleration region, see also recent review papers on the particle acceleration in solar flares \citep{Zharkova_etal_2011, Petrosian_2012}.

Nevertheless, the standard paradigm of the stochastic acceleration is not apparently the whole story since it cannot easily predict the acceleration efficiency leading to the observed variety of the thermal-to-nonthermal energy partitions from 'heating without acceleration' \citep[{at least, at the early stage of some flares, see} e.g.,][]{Battaglia_etal_2009} to 'acceleration without heating' \citep{Fl_etal_2011}, which, perhaps, requires a specific mechanism capable of dissipating the flare energy into the plasma heating and at the same time extracting a fraction of particles from the thermal pool and supplying them to the stochastic acceleration process, {see \cite{Altyntsev_etal_2012} for more detail}. Furthermore,
it is at odds, e.g., with particle acceleration in form of beams\footnote{Although the beams can well be formed outside the acceleration region due to transport effects \citep[e.g.,][]{Reid_etal_2011, Vilmer_2012}}  \citep[detected via radio type III bursts or microwave continuum, see, e.g.,][]{Altyntsev_etal_2007, Altyntsev_etal_2008}. {It is not easily consistent with} huge enhancement of some ion abundances including $^3He$ and ultra heavy ions {\citep[see, however,][]{Liu_etal_2006}} observed in solar energetic particle (SEP) events produced from impulsive solar flares \citep{Koch_Koch_1984,Reames_1999,Mason_2007}, and also indirectly implied by the enhanced positron production at the flare site \citep{Kozlovsky_etal_2004}. In addition, \citet{Share_Murphy_1998} analyzed various gamma-ray lines produced by different accelerated ions at the flare site and found that the $He/p$ ratio was noticeably enhanced (by a factor of 5) relative to the standard photospheric value.


Then, the RHESSI discovery of distinctly different spatial locations for hard X-ray emission produced by accelerated electrons and gamma emission produced by ions in the same flare \citep{Lin_etal_2003, Hurford_etal_2006, Vilmer_etal_2011} puts further constraints on particle acceleration in flares. The first of the events, 23 July 2003, where this separation was observed, displayed a complex source structure with the HXR (gamma-) source associated with shorter (larger) post-flare loops, which has been interpreted in terms of resonant stochastic acceleration by cascading MHD turbulence \citep{Emslie_etal_2004}. Since then, a few more strong gamma-ray flares have been studied.  Although no statistically significant set of flares with gamma-ray imaging has been accumulated \citep{Vilmer_etal_2011}, the several available events display a variety of spatial relationships with a significant spatial separation between the electron and ion emission sites.  In the cases (two of five available events) where the ion emission site coincided with a HXR footpoint (which seems at odds to \citet{Emslie_etal_2004} model), it was a \textit{weaker} HXR footpoint, indicating that the accelerated electrons and ions preferably precipitate into the \textit{opposite} footpoints of a flaring magnetic loop or loop system; something that could also happen in the other events, although the magnetic connectivity between sources is less clear in those cases.

These  observed properties of particle acceleration in flares imply that an important ingredient is \textit{missing} from our understanding of the flare acceleration process commonly ascribed to the stochastic acceleration mechanism \citep{Miller_etal_1997, Aschw_2002, Emslie_etal_2004, Zharkova_etal_2011, Petrosian_2012}. Indeed, within the concept of \textit{stochastic acceleration}, in contrast to \textit{direct acceleration} by a DC electric field, the particles  being scattered by stochastically turbulent magnetic fields move diffusively (at least, inside the acceleration region) and, according to the Fisk law, form a particle flux in the direction opposite to their density gradient, which is supposed to be independent on the particle charge sign. Some asymmetry between two opposite footpoint emission is expected due to asymmetry of the magnetic loop; however, this asymmetry must be the same for both protons and electrons, not the opposite as observed. Having the protons and electrons precipitating into the opposite footpoints would be natural in presence of a DC electric field. The DC electric fields can accelerate the charged particles directly but to a modest energy ($\lesssim100$~keV) only  \citep[e.g.,][and references therein]{Holman85, Aschw_2002, Zharkova_etal_2011}; thus, an acceleration process combining features of both stochastic acceleration and acceleration in DC electric field is called for.

We are not going, however, to arbitrarily postulate these two highly distinct ingredients---a regular DC electric field and random turbulent fields needed for  stochastic acceleration. Instead, we envision a model where the appearance of regular electric fields is an intrinsic property of the very same turbulence that gives rise to the stochastic acceleration---a helical turbulence model---in which the effect of helicity does not substitute but complement the more well-known stochastic acceleration process. We demonstrate below that the helicity of the turbulence created in flaring loops is an unavoidable direct consequence of the magnetic field nonpotentiality at the flare site, which is itself needed to drive the flare energy release. Furthermore, we show that the turbulence kinetic helicity in the corona is coherently built up from non-zero helicity of individual linear MHD waves forming the turbulence, which is linked to the current helicity and so can reliably be estimated using measurements of the current helicity at the photosphere. Stated another way, the helicity is an unavoidable property of the turbulence created in the nonpotential, twisted flaring loops; account of the turbulence helicity effect on the stochastic acceleration offers, as we show below, a nice natural way of addressing the mentioned outstanding questions about particle acceleration in flares including the thermal-to-nonthermal energy partition, the electron/ion spatial separation, formation of electron beams, and significant $^3He$ enrichment.

\section{Particle Acceleration by Helical Turbulence} 
\label{S_acc_H_turb}

Theory of particle acceleration by helical turbulence was proposed long ago by \citet{Kichat_1983} and then developed by \citet[][and references therein]{Fedorov_etal_1992, Fedorov_Stehlik_2010, Fedorov_2011} in relation to Galactic cosmic ray acceleration problem. Since it has not been discussed yet in the solar context, we outline below the essence of the theory.
Let us denote the exact (fluctuating) distribution function of the fast particles of a given kind as $f(\vr,\vp,t)$; apparently, it satisfies the collisionless kinetic equation\footnote{The collisional integral is discarded assuming the acceleration dominates over the Coulomb losses; we return to the role of Coulomb collision, that is important for thermal particles, later in \S~\ref{S_Dreicer}}:

\begin{equation}\label{9_5}
\frac{\partial{f}}{\partial{t}}+\vv\cdot\frac{\partial{f}}{\partial\mathbf{r}}+\vcF\cdot\frac{\partial{f}}{\partial\mathbf{p}}=0.
\end{equation}
The force
\begin{equation}\label{9_6}
 \vcF=e\vE+\frac{e}{c}\vv\times\vB=\frac{e}{c}(\vv-\vu)\times\vB
\end{equation}
contains both electric and magnetic components, where we employ the high plasma conductivity providing $\vE=-\vu\times\vB/c$.

Let us consider the Lorenz force structure in the presence of a regular magnetic field $\vB_0(\vr)$ with a turbulent wave spectrum superimposed. The turbulence can be characterized by a random magnetic field vector $\vb$ and a random vector of the fluid velocity $\vu$ (the fluid is adopted to be at rest so no regular flow velocity is present). Thus, substituting $\vB=\vB_0+\vb$ and $\vu$ into Eqn~(\ref{9_6}) we obtain

\begin{equation}\label{9_6.2}
 \vcF=\frac{e}{c}\left[\vv\times\vB_0  +\vv\times\vb -\vu\times\vB_0 -\vu\times\vb\right].
\end{equation}
The first term describes the particle gyration in the regular magnetic field $\vB_0$, the second one---the particle angular scattering by magnetic irregularities $\vb$, while the third one---the standard stochastic (second order Fermi) acceleration by the turbulent electric field $\vE=-\vu\times\vB_0/c$. Remarkably, the second and third terms have zero means; thus, they result in the second order diffusion in the momentum space---isotropization and stochastic acceleration, respectively. The fourth term, also representing an electric field $-\vu\times\vb/c$ is, however, different from the third one because it can have a non-zero mean $\left<\vu\times\vb\right>\neq0$, which will represent a regular electric field capable of directly accelerating the charged particles. Importantly,  having this nonzero regular electric field constrains the turbulence in a certain way. Indeed, based on the symmetry consideration, the mean electric field  $\widetilde{\vE}=\left<\vu\times\vb\right>/c$ must be directed along the mean magnetic field,
\begin{equation}\label{Eq_Emean_est}
\widetilde{\vE}=\widetilde{p}\vB_0,
\end{equation}
and given that the electric field is polar, while the magnetic field is axial vector, we conclude that the coefficient $\widetilde{p}$ must be a pseudoscalar. Stated another way, only such turbulence whose measures intrinsically include a pseudoscalar can result in a nonzero regular electric field $c\widetilde{\vE}=\left<\vu\times\vb\right>$. In what follows we show that this is the case of \textit{helical turbulence}.

To consider this effect quantitatively let us
write down Eqn. (\ref{9_5}) in the form:

\begin{equation}\label{9_7}
 \frac{\partial{f}}{\partial{t}}+\vv\cdot\frac{\partial{f}}{\partial\mathbf{r}}+\vB\cdot\widehat{\vD}\,f=0,
\end{equation}
where
\begin{equation}\label{9_8}
\widehat{\vD}=\frac{e}{c}(\vv-\vu)\times\frac{\partial}{\partial\mathbf{p}}
\end{equation}
is the operator that changes the value and direction of the particle momentum.
As has been said above we adopt $\vu$ to be a purely
random vector with zero mean  $\langle\vu\rangle=0$.

Let us average Eqn (\ref{9_7}) over the turbulence  ensemble as is commonly done to address the stochastic acceleration problem. For simplicity, we assume that the average distribution function does not depend  on coordinates, so that
\begin{equation}\label{10.40}
  {\langle}f(\vr,\vp,t)\rangle=F(\vp,t), \qquad   f=F+\delta f(\vr,\vp,t)  .
\end{equation}
After averaging we obtain the equation
\begin{equation}\label{10.41}
  \frac{\partial{F}}{\partial{t}}+\langle\vcF\rangle\cdot\frac{\partial{F}}{\partial\vp}=\frac{v}{2\Lambda(p)}\widehat{\vcO}^2\,F\,+\,\widehat{S}_aF,
\end{equation}
where each term is primarily linked with a certain component of the Lorenz force in Eqn~(\ref{9_6.2}): the first term at the rhs originates from averaging $\langle \vb \delta f(\vr,\vp,t)\rangle$ and describes angular scattering of the particles by magnetic irregularities, which does not change the particle energy, while the second term originating from the electric field averaging, $\langle \vu \delta f(\vr,\vp,t)\rangle$, describes the standard stochastic acceleration by turbulence.
%
%
The angular scattering term is expressed via the particle mfp $\Lambda(p)$ that is set up by the turbulent magnetic fluctuation spectrum. To be specific, we adopt the power-law spectrum
\begin{equation}\label{9_36}
 \langle b^2\rangle_k dk=\frac{(\nu-1)k_c^{\nu-1}\langle
 b^2\rangle}{k^\nu}dk,\q\nu>1, \q k_c < k < 2\pi/l_{\min},
\end{equation}
where  $b=\sqrt{\langle b^2\rangle}$  is the total rms turbulent magnetic field so that $\langle b^2\rangle=\int \langle b^2\rangle_k dk$, $\nu$ is the turbulence power-law spectral index, $L_c=2\pi/k_c$ is the main (largest, energy-containing), while $l_{\min}$ is the minimal scale of the turbulence, which can decrease at the course of the turbulence cascading in contrast to the main scale $L_c$, which remains roughly constant. Then, the mfp
$\Lambda$ is  given by \citep[\S~15.3, pp 207--211, in][Eqns~78,~13, respectively]{Toptygin_1985, Schlickeiser_1989, Petrosian92}
\begin{equation}\label{9_37}
 \Lambda\approx\frac{B_0^2}{\langle{b}^2\rangle}\left(\frac{R_0}{L_c}\right)^{2-\nu}L_c \propto p^{2-\nu},
\end{equation}
$\widehat{\vcO}=\vv\times\frac{\pa}{\pa\vv}$ is an angle-alternating operator, and
the operator $\widehat{S}_a$ includes all acceleration effects  other than that produced  by the nonzero mean electric field $\widetilde{\vE}=\left<\vu\times\vb\right>/c$ if present. 

Below we concentrate on this new  effect  described by the term proportional to
$\langle\vcF\rangle$, qualitatively distinct from what is commonly included in the standard Fermi acceleration. As has been noticed based on qualitative consideration of the Lorenz force in form (\ref{9_6.2}), if $\vu$ and $\vb$ at a given point are correlated and not parallel to each other then $\left<\vu\times\vb\right>\neq0$, which yields
\begin{equation}\label{10.42}
  \langle\vcF\rangle=\frac{e}{c}\vv\times\vB_0-\frac{e}{c}\langle\vu\times\vb\rangle,
\end{equation}
where the last term represents an electric force
$\langle\vcF_h\rangle$ acting on the particles.

In what follows we employ a discovery made by
Parker, Steenbek and Krause \cite[see][]{Vainshtein_Zeldovich_1972, Parker_1979, Vajnshtejn_etal_1980, KR80}
within the turbulent dynamo theory that
large-scale fields can be produced by a turbulence lacking mirror
symmetry (helical turbulence). This
ability to amplify the large-scale \textit{magnetic} field is the main driver
of studying the helical turbulence in the astrophysical context, although for the particle acceleration problem addressed here we are primarily interested in the accompanying large-scale \textit{electric} field, which is an unavoidable byproduct of the turbulence kinetic helicity; see Appendix~II.

The correlation tensor of this turbulence contains both polar and axial
vectors. Perhaps the simplest example of such structure represents
the correlation tensor of statistically uniform gyrotropic
incompressible fluid:
\begin{equation}\label{5.12}
\widetilde{U}_{\mu\nu}(\vk,t)=A(k,t)\left(\delta_{\mu\nu}-\frac{k_\mu{k_\nu}}{k^2}\right)-iP(k,t)e_{\mu\nu\lambda}k_\lambda.
\end{equation}
Importantly, the function $P(k,t)$ here must be a pseudoscalar (recall, the presence of a pseudoscalar is a necessary condition for having a nonzero mean electric field $\widetilde{\vE}$),
i.e., it changes the sign at the coordinate inversion.
The turbulence helicity can be quantified by the following integral parameter
\begin{equation}
\label{5.13}%
\alpha\,=\,-\frac{1}{3}\int_0^\infty\langle\vu(\vr,t+\tau)\cdot\nabla\times\vu(\vr,t)\rangle{d\tau}=
$$$$
-\frac{\tau_c}{3}\langle\vu(\vr,t)\cdot\nabla\times\vu(\vr,t)\rangle=-\frac{\tau_c}{3\pi^2}\int_0^\infty{k^4}P(k,0)dk.
\end{equation}%
This measure, $\alpha$, is called the \textit{kinetic helicity parameter}, which is defined  by the second term in Eqn (\ref{5.12}) only.

Apparently, a nonzero helicity is only possible for the systems where a pseudoscalar 
can be built from  available physical
parameters; for example it is the case of the dot product of a polar vector (e.g., gradient of pressure or density, which requires plasma inhomogeneity) and an axial vector (e.g., the magnetic field).  In particular, in the magnetized coronal plasma the helicity can well be driven by the twist of the nonpotential field lines, where the polar vector of the electric current density $\vj$ is available.
It is, therefore,
important  that the mean magnetic field is itself \textit{twisted}, so that
\begin{equation}\label{10.39}
    \vB=\vB_0+\vb,\q\langle\vB\rangle=\vB_0\neq0,\q \nabla\times\vB_0\neq0,\q\langle\vb\rangle=0,
\end{equation}
where $\vB_0(\vr)$ is a mean regular \textit{nonpotential} magnetic field in the acceleration region and $\vb$ is the fluctuating turbulent magnetic field.

As soon as the helicity parameter has been specified (we return to estimating it later, see Appendixes), the averaged electric force
is straightforwardly calculated within the turbulent dynamo theory \citep{Vainshtein_Zeldovich_1972, Parker_1979,KR80, Kichat_1983} and neglecting the dissipation
(and magnetic field nonuniformity) it has the form agreeing with expectation (\ref{Eq_Emean_est})

\begin{equation}\label{10.43}
\langle\vcF_h\rangle=-\frac{e}{c}\langle\vu\times\vb\rangle=e\vE_h=-\frac{e}{c}\alpha\vB_0,
\end{equation}
which depends on the turbulence helicity only, while does not depend on other plasma properties, e.g., on the plasma beta.
Hence, Eqn (\ref{10.41}) for the particle distribution function receives the form

\begin{equation}\label{10.44}
\frac{{\partial}F}{\partial{t}}+\frac{ec}{\cE}\vB_0\cdot\widehat{\vcO}F-\frac{e\alpha}{c}\vB_0\cdot\frac{\partial{F}}{\partial\vp}=\frac{v}{2\Lambda(p)}\widehat{\vcO}^2\,F,
\end{equation}
where for now we have neglected the contribution from the term
$\widehat{S}_aF$, since we are interested in 
the contribution coming from
the large-scale electric field $(-\alpha/c)\vB_0$; however, we will compare efficiencies of these two acceleration processes later.

The regime of particle acceleration including the acceleration rate in the DC electric field depends on the particle transport regime, which is controlled by the
particle mfp $\Lambda(p)$. At a relatively small scale, $l\ll \Lambda$, the particle guiding center moves almost rectilinearly along the mean magnetic field and the energy gain of a nonrelativistic particle is  proportional to $t^2$, i.e.,  the energy gain is fast. Apparently, at this stage the particles can form a beam and quickly gain energy of the order of\footnote{This estimate is valid if the source size $L$ is larger than the particle mfp; otherwise the estimate apparently reads $eZE_h L$.} $eZE_h\Lambda$, where $eZ$ is the charge of the particle. At larger scales, $l\gg \Lambda$, the particle experiences many scatterings and so moves diffusively back and forth along the magnetic (and, having the same direction, DC electric) field. Accordingly, the particle gains energy from the DC electric field when it moves one direction, while loses it when  moves the opposite direction; thus, the particle energy changes much slower with time, than at the small scale regime. Even though the particles move diffusively, the process of their energy gain from the regular DC electric field is different compared with that from stochastic electric fields during standard stochastic acceleration process. Indeed, consider the energy change of a particle, which comes back to its original location after it has randomly walked for a while. In the case of purely stochastic electric fields this change is also stochastic---it can be positive or negative, being positive \textit{on average}. In contrast, in the case of purely regular DC electric field this energy change is zero equivalently. Here, the mean \textit{positive} energy gain \textit{in time} is provided by regular unidirectional drift of the particles, driven by this DC electric field, superimposed on the particle diffusion.
Stated another way, regardless of the particle transport regime the particle energy gain in a DC electric field is a \textit{regular function} of its original and final \textit{spatial locations}; for the small scales (small source size) it also a regular function of time, while for large scales (large source size) it is a \textit{random function of time}; thus, individual particles can either gain or lose their energy depending on their individual paths, while gaining the energy with time \textit{on average} even though the accelerating field is a regular DC electric field.


Since the solution for the small scale regime, $l\ll \Lambda$, is trivial, we consider the other regime, $l\gg \Lambda$, in more detail.
In a general case kinetic equation (\ref{10.44})  describes anisotropic particle distributions. However, if the departure from the isotropy is small, which is reasonable to expect within the considered transport regime because of the efficient angular scattering of the particles by the very same turbulence described by the rhs of Eqn (\ref{10.44}), this equation can be further simplified by using the diffusion approximation (we emphasize that this will also be very convenient for the purpose of further comparison of this mechanism with the temporarily discarded statistical Fermi acceleration) if the source size $L$ is much larger than the particle mfp  $L\gg\Lambda$. Assuming this to be the case\footnote{For example, adopting a modest turbulence level of $b^2/B^2\sim10^{-4}$, $L_c\sim2\cdot10^7$~cm, $\nu=1.5$, and $R_0\sim0.2$~cm, Eqn~(\ref{9_37}) yields $\Lambda\sim2\cdot10^7$~cm for electrons, which is indeed much smaller than the size of acceleration region, $\sim10^9$~cm, determined from observations \citep[e.g.,][]{Fl_etal_2011}.}, we adopt a weakly anisotropic distribution function to have the form:

\begin{equation}\label{9_28}
F(\vr,\vp,t)=\frac{1}{4\pi}\left[N(\vr,p,t)+\frac{3}{v^2}\,\vv\cdot\vJ(\vr,p,t)\right],
\end{equation}
where
\begin{equation}\label{9_29}
 N(\vr,p,t)={\int}F(\vr,\vp,t)d\Omega_p,\q
 $$$$
 \vJ(\vr,p,t)=\int{\vv}F(\vr,\vp,t)d\Omega_p
\end{equation}
are, respectively, the isotropic part of the distribution function and the flux density of the particles with a given energy; weak anisotropy implies
$J{\ll}vN$.

Substituting Eqn (\ref{9_28}) into Eqn (\ref{10.44}), we
obtain the set of coupled equations
\begin{subequations}
\label{10.45}
\begin{equation}\label{10.45a}
\frac{\partial{N}}{\partial{t}}=\frac{e\alpha}{cpv}\vB_0\cdot\left[\left(1+\frac{v^2}{c^2}\right)\vJ+p\frac{\partial\vJ}{\partial{p}}\right],
\end{equation}
\begin{equation}\label{10.45b}
\frac{1}{v}\frac{\partial\vJ}{\partial{t}}+\frac{ec}{v\cE}\vB_0\times\vJ-\frac{e\alpha}{3c}\frac{\partial{N}}{\partial{p}}\vB_0=-\frac{1}{\Lambda}\vJ
\end{equation}
\end{subequations}
for the isotropic part of the particle distribution function $N$ and the flux density $\vJ$.
Note that the direction of the flux density depends on the charge sign, implying opposite flux directions for the protons and electrons, respectively.
Remarkably, it also depends on the $\alpha$ sign. For a flaring loop in the solar corona this means that for a given magnetic polarity of the loop footpoints at the photosphere level, the electric field direction (and, accordingly, directions of positive and negative charge flows) differs, respectively,  for the left and right twists of the magnetic flux tube---a prediction, which can be tested observationally.

Let us eliminate $\vJ$ from (\ref{10.45a})  neglecting the term
$(1/v)\pa\vJ/\partial t$ in Eqn~(\ref{10.45b}). We then arrive at the diffusion equation {for the isotropic part $N$ of the weakly anisotropic distribution function $F$, Eqn~(\ref{9_28}),} in the momentum space \citep{Kichat_1983},
\begin{equation}\label{10.46}
\frac{\partial{N}}{\partial{t}}=\frac{1}{p^2}\frac{\partial}{\partial{p}}p^2D_h\frac{\partial{N}}{\partial{p}},
\end{equation}
where
\begin{equation}\label{10.47}
D_h=\frac{\alpha^2\Lambda{p^2}}{3vR_0^2}\propto \frac{p^{2-\nu}}{v}
\end{equation}
is the diffusion coefficient which describes the particle
acceleration by the helical turbulence, and $R_0=cp/eB_0$ is the Larmor radius. Eqn~(\ref{10.46}) ignores finite size of the acceleration region and so permits acceleration up to arbitrary high energy; in fact, for a flux tube with a length $L$ the maximum energy can apparently be estimated as $\cE_{\max}\approx 3eE_h L$.

In spite of the presence of the regular electric field along the regular magnetic field, we have arrived at a diffusive form (in the momentum space) of equation describing particle acceleration by the helical turbulence. As has been noticed above, this happens because of efficient particle angular scattering by the turbulence, which makes local particle motion  back and forth relative to the accelerating electric field almost equally probable, although the probability to gain energy remains larger than to lose energy. Accordingly, though the regular electric field enhances explicitly only the parallel particle momentum, the angular scattering transfers it to the transverse momentum, so all momentum components rise proportionally roughly preserving the distribution isotropy.

In fact, the (discarded earlier) second order Fermi-acceleration process  does contribute to the right-hand side of Eqn (\ref{10.46}) in addition to the acceleration due to turbulence helicity. The Fermi-acceleration diffusion  coefficient is given by \citep{Toptygin_1985, Fedorov_etal_1992}
\begin{equation}\label{10.48}
D_F=\frac{\langle {u^2}\rangle{p^2}}{3v\Lambda}\propto \frac{p^{\nu}}{v}.
\end{equation}
Thus  the helical part of the velocity correlation tensor gives a dominant contribution to the acceleration if
\begin{equation}\label{10.49}
\frac{D_h}{D_F}=\frac{\alpha^2}{\langle{u^2}\rangle}\frac{\Lambda^2}{R_0^2}\gg1
\end{equation}
{in agreement with Fedorov et al., 1992, Eqn~(95).}
For cosmic plasmas, as a rule, $\Lambda\gg{R_0}$. The
helical parameter $\delta=\alpha^2/\langle{u^2}\rangle$ is generally unknown and hard to reliably estimate; 
it
seems to be rather small for most of the astrophysical objects. Below we make the case that the turbulence helicity can, however, be firmly calculated when it is driven by twist of the mean magnetic field in the acceleration region, which is likely relevant to particle acceleration in solar flares.

\section{Kinetic Helicity of MHD turbulence in Nonpotential Flaring Loops}

The charged particle  acceleration in solar flares occurs in a relatively tenuous corona, where  no direct measurement or estimate of the kinetic helicity is available. To get a qualitative idea of possible origin of the helical turbulence in the solar flaring loops and the order of magnitude of the helicity parameter\footnote{A formal quantitative derivation based on the MHD equation set is given in  Appendix~I.} $\alpha$ we consider an MHD turbulence composed of random ensemble of Alfv\'en waves with a broad distribution over the wavelength, and, perhaps, containing other linear eigen-modes of the magnetized plasma. A single Alfv\'en wave represents in fact an oscillation of the corresponding magnetic filed line (or a magnetic flux tube); in the presence of the wave ensemble this field line will experience many independent oscillations linearly superimposed on each other until their amplitudes remain small. If the mean magnetic field is uniform then no preferable twist is expected in the turbulence composed of these linear wave-modes. If, however, the original field line is itself twisted then its oscillations will inherit this twist, resulting in a nonzero helicity in the general case of the MHD turbulence generated on top of a nonpotential twisted mean magnetic field, which is, in particular, the case of the solar coronal flaring loops.

To obtain a quantitative estimate we
express the helicity parameter in the form $\alpha=-\tau_c \langle h_k\rangle/3$, where $h_k$ is the kinetic helicity density defined as $h_k=\vu(\vr,t)\cdot\nabla\times\vu(\vr,t)$, sf Eqn (\ref{5.13}).  The advantage of this representation, as we are going to demonstrate below, is that in the case of the solar corona we can reliably  estimate the \textit{kinetic} helicity density using observational data on the \textit{current} helicity density {(which can be better quantified for the solar corona than the magnetic helicity used by Fedorov et al. 1992 although one can be expressed via the other)} defined as

\begin{equation}\label{Eq_Gyro_Turb_accel_h_c}
h_c=\vB(\vr,t)\cdot\nabla\times\vB(\vr,t)=\alpha_{FFF}B^2,
\end{equation}
where the second equality is written for a force-free field (FFF),  which implies a direct link between the force-free parameter $\alpha_{FFF}$ and the current helicity density $h_c$ for the case when the magnetic field satisfies the force-free conditions. Although none of these two parameters can be directly measured in the corona, we make use of $\alpha_{FFF}$ conservation in the force-free field along any given field line ($\vB\cdot\nabla\alpha_{FFF}\equiv0$) and use the photospheric current helicity measurements (or a corresponding nonlinear FFF extrapolation, see, e.g., \citet{Jing_etal_2012} and references therein), which implies that we can have a reliable estimate of the current helicity in the corona.

Let us now estimate how the turbulence kinetic helicity is related to the current helicity. For the linear MHD modes, the fluid velocity is approximately proportional to the magnetic field in the waves,
\begin{equation}\label{Eq_u_b_MHD}
\vu\approx\mp{v_A}\vb/B,
\end{equation}
thus, assuming $\rho=\const$ for simplicity, we obtain
\begin{equation}\label{Eq_Gyro_Turb_accel_h_k}
h_k=\vu(\vr,t)\cdot\nabla\times\vu(\vr,t)=\frac{v_A^2}{B^2}\vb(\vr,t)\cdot\nabla\times\vb(\vr,t).
\end{equation}
The last needed step is to estimate the current helicity component related to only the turbulent magnetic field,
$\hat{h}_c=\vb(\vr,t)\cdot\nabla\times \vb(\vr,t)$, which can be done based on the full current helicity $h_c$, Eqn (\ref{Eq_Gyro_Turb_accel_h_c}).

To obtain this link let us write down Eqn~(\ref{Eq_Gyro_Turb_accel_h_c}) for a field line distorted by an Alfv\'en wave given that even in the presence of the wave the force free condition remains on average valid and take into consideration the $\alpha_{FFF}$ conservation along this new (distorted) field line,
\begin{equation}\label{Eq_Gyro_Turb_h_link}
\langle(\vB_0(\vr,t)+\vb(\vr,t))\cdot\nabla\times(\vB_0(\vr,t)+\vb(\vr,t))\rangle=
$$$$
\alpha_{FFF}\langle(\vB_0(\vr,t)+\vb(\vr,t))^2\rangle.
\end{equation}
After averaging Eqn (\ref{Eq_Gyro_Turb_h_link}), all linear over $\vb$ terms drop out and then, subtracting the original equality $\vB_0\cdot\nabla\times\vB_0=\alpha_{FFF}\vB_0^2$ from Eqn (\ref{Eq_Gyro_Turb_h_link}), we obtain
\begin{equation}\label{Eq_Gyro_Turb_accel_hat_h_k}
\langle\hat{h}_c\rangle\equiv\langle\vb(\vr,t)\cdot\nabla\times\vb(\vr,t)\rangle=\alpha_{FFF}\langle b^2\rangle,
\end{equation}
so the kinetic helicity density takes the form (sf Eqn~(\ref{5t} in Appendix~I):
\begin{equation}\label{Eq_Gyro_Turb_accel_h_k_FFF}
\langle{h_k}\rangle\approx\alpha_{FFF}v_A^2\frac{\langle{b^2}\rangle}{B^2},
\end{equation}
where $\langle{b^2}\rangle$ is related to the entire turbulent magnetic field because all waves composing the turbulence are statistically independent so their contributions add up incoherently.
Now, estimating $\tau_c$ as $L_c/v_A$, where $L_c$ is the main scale of the turbulence, see Eqn~(\ref{9_36}), we obtain, in agreement with Eqn~(\ref{13.t}) of Appendix~II, the required kinetic  helicity parameter
\begin{equation}
\label{Eq_Gyro_Turb_accel_heli_alpha_corona}
\alpha\approx-\frac{\tau_c\langle{h_k}\rangle}{3}\approx-\alpha_{FFF}L_cv_A\frac{\langle{b^2}\rangle}{3B^2}.
\end{equation}

This estimate of the helicity parameter enables us of estimating relative efficiency of  the 'helical' acceleration and the standard stochastic (Fermi) acceleration given by Eqn (\ref{10.49}), which, with the account of Eqns (\ref{Eq_Gyro_Turb_accel_heli_alpha_corona}) and (\ref{Eq_u_b_MHD}), yields

\begin{equation}\label{10.49_estimate}
\frac{D_h}{D_F}\sim\frac{\alpha_{FFF}^2L_c^2b^2}{B^2}\frac{\Lambda^2}{R_0^2}\sim\alpha_{FFF}^2L_c^2 \frac{B^2}{b^2}\left(\frac{L_c}{R_0}\right)^{2(\nu-1)}
 \propto p^{-2(\nu-1)},
\end{equation}
where in the second equality we use Eqn (\ref{9_37}) to estimate the particle mfp formed by the same turbulence.
It is transparent from here that the helical acceleration is relatively more important for (i) more nonpotential loops (larger $\alpha_{FFF}$),
(ii) weaker turbulence or stronger mean magnetic field (smaller $b^2/B^2$ ratio), and (iii) lower energy (smaller Larmor radius $R_0$).

We now in the position to make estimates for various numeric inputs of the relevant parameters.
For example, substituting into Eqn~(\ref{10.49_estimate})  modest values of $\alpha_{FFF}\sim10^{-10}$~cm$^{-1}$ \citep{Abramenko_etal_1996, Abramenko_etal_1997, Longcope_etal_1998},
$R_0\sim1$ cm,\footnote{For example, for $B=100$~G and $T=1$~MK, the thermal gyroradius of electron is $\sim0.2$~cm, while of proton is $\sim10$~cm.}  $L_c\sim10^7$ cm, $\nu=1.5$, and $(b/B_0)^2\sim10^{-4}$,
we find $D_h/D_F\sim10^5$.
Note that in flare-productive active regions the $\alpha_{FFF}$ parameter can be much larger than adopted above: from the nonlinear force free field modeling of  active region 11158 (February 2011) \cite{Jing_etal_2012} reported the mean and maximum values to be $|\alpha_{FFF}|=5.2\cdot10^{-9}$~cm$^{-1}$ and $|\alpha_{FFF}|=5.8\cdot10^{-7}$~cm$^{-1}$, respectively. Consider now an extreme case of a strong turbulence $(b/B_0)^2\sim1$, which, presumably, happens in a highly twisted flaring loop. {The twist of a quasistationary loop, cannot, however, be arbitrarily large since the loop is known to become unstable \citep[e.g.,][]{Browning_Priest_1983} if the number of twists, $N_{tw}$, is large,  $N_{tw}\gtrsim 3$ \citep[][p. 269-270]{Aschwanden_2006}. For a loop with the length $L\sim 10^9$~cm this translates to $\alpha_{FFF}\lesssim \alpha_{FFF,cr}\approx 4\pi N_{tw}/L\approx3\cdot10^{-8}$~cm$^{-1}$ \citep[][p. 215]{Aschwanden_2006}. For $\alpha_{FFF}\sim \alpha_{FFF,cr}$ we find $D_h/D_F>1$ for $L_c>10^5$~cm, which is likely to be fulfilled for any realistic turbulence model,
i.e., acceleration by helical turbulence can be up to
a few orders of magnitude more efficient for low-energy particles in the solar corona than the standard stochastic acceleration by the same turbulence. \cite{Jing_etal_2012} reported even one order of magnitude higher values $|\alpha_{FFF}|=5.8\cdot10^{-7}$~cm$^{-1}$ implying even stronger acceleration, which, however, from the loop stability perspective can only happen in rather short loops, $L\lesssim 10^8$~cm.}
Note, that in addition to the overall limitation of the helical acceleration energy gain by the value $\sim 3eZE_h L$, which cannot exceed 100~keV for any reasonable input parameters, the particle Larmor radius grows with energy, so the relative
efficiency of the acceleration by helical turbulence drops and
the standard stochastic (Fermi) acceleration can start to dominate even earlier  at some energy $<3eZE_h L$ in case of a reasonably strong MHD turbulence.
Thus,  all known advantages of the stochastic acceleration at  \textit{high energies} remain unaffected by the turbulence helicity.


However, the helical acceleration is efficient
at low energies and for a weaker turbulence and so can, perhaps, more efficiently, than the standard stochastic acceleration process, extract charged particles from the thermal pool and, thus,  form a particle seed population from which the particles are then picked up by the main, presumably stochastic, acceleration process.
Indeed, the turbulence helicity builds up a large-scale regular electric field which can form a runaway particle population. The runaway electrons represent collimated beams, which are supposed to reveal themselves via radio type III bursts \citep{Aschwanden_etal_1990, Vlahos_Raoult_1995} and via sub-second fluctuations of microwave \citep{Altyntsev_etal_2008} or X-ray \citep{Kiplinger_etal_1983, Aschwanden_etal_1996} emissions. In particular, if the turbulence consists of many turbulent 'cells', where the mean magnetic field has different directions relative to the line of sight, then the directions of the electron beams will also be different in agreement with observations in the microwave range \citep{Altyntsev_etal_2007}. However, at the thermal energy range the Coulomb collisions are important, whose effect we investigate in the next section.

\section{Implications for Particle Seed Population}
\label{S_Dreicer}

{In the considered model of acceleration by helical turbulence, three main ingredients are responsible for the particle distribution function formation---Coulomb collisions, stochastic turbulent electric and magnetic fields, and the mean DC electric field.  Clear division of the particles onto the thermal and nonthermal components in the acceleration region is straightforward when binary collisions is the dominant process for the thermal particles (setting the thermal Maxwellian distribution), while less important or entirely negligible for high-energy particles (allowing a nonthermal, non-Maxwellian distribution). Therefore, to transfer a particle from the thermal pool to the nonthermal component requires a process (commonly called 'injection') capable of overcoming the Coulomb losses. We have already established that the role of the helical turbulence-induced mean DC field is often more important compared with the stochastic fields at low energies; thus, the injected particle population is supposed to be formed in a tradeoff between the (accelerating) DC electric field and (decelerating) Coulomb collisions, where the latter can be quantified by the Dreicer field.
} {Note that in addition to the Coulomb collisions, the collisionless angular scattering by the turbulent waves can be present even for the thermal and suprathermal particles. However, since this scattering does not imply any energy loss, it is inessential to the considered here injection threshold and efficiency and so will be neglected, even though it can affect the exact dynamics of this process. To be specific, we adopt that there are no turbulent waves capable of resonantly interact with the thermal and slightly suprathermal particles because the corresponding linear modes have a strong damping; thus, the thermal particle transport is mediated by the Coulomb collisions.  }
Let us estimate how big the regular electric field induced by the helical turbulence can be compared with the electron Dreicer field.
Substituting Eqn (\ref{Eq_Gyro_Turb_accel_heli_alpha_corona}) into  Eqn (\ref{10.43}) we obtain the electric field in the form

\begin{equation}
\label{Eq_E_h}
    E_h= \left(\begin{array}{c}2.4\cdot10^{-7}\\7.3\cdot10^{-5}\end{array}\right)\left(\frac{B}{100~{\rm G}}\right)^2\left(\frac{n_e}{10^{10}~{\rm cm}^{-3}}\right)^{-1/2}$$$$
    \left(\frac{\alpha_{FFF}}{10^{-10}~{\rm cm}^{-1}}\right)\left(\frac{L_c}{10^8~{\rm cm}}\right)
    \left(\frac{\langle{b^2}\rangle/B^2}{10^{-4}}\right)\left(\begin{array}{c}{\rm{statvolt/cm}}\\{\rm{V/cm}}\end{array}\right)
    ,
\end{equation}
which, for reasonably modest assumptions on the acceleration region parameters yields%

\begin{equation}
\label{Eq_Ereg_h_est}
E_h\sim 10^{-5}-10^{-4} \ \ {\rm{V/cm}}.
\end{equation}
Efficiency of particle acceleration by a DC electric field is specified by its ratio to the Dreicer field, whose value is defined by the particle Coulomb collisions. In particular, the electron Dreicer field  is

\begin{equation}
\label{Eq_Accel_Dreicer_field}
    E_{De}= \frac{e\ln\Lambda_C}{r_d^2}=
    \left(\begin{array}{c}  2\cdot 10^{-7} \\
                    6\cdot 10^{-5}
                    \end{array} \right)
     \left(\frac{n_e}{10^{10}~{\rm cm}^{-3}}\right)
     $$$$
     \left(\frac{T}{10^7~{\rm K}}\right)^{-1}\left(\frac{\ln\Lambda_C}{20}\right)~~\left(\begin{array}{c}  {\rm statvolt/cm} \\
                    {\rm V/cm}
                    \end{array} \right),
\end{equation}
%
therefore, in our example the induced electric field, Eqn~(\ref{Eq_Ereg_h_est}), is slightly sub-Dreicer, although it can exceed the Dreicer field, e.g., for more tenuous plasma and/or for stronger turbulence. The sub-Dreicer field  creates beams of the runaway electrons with the velocity exceeding the critical velocity, $v_c\sim{v_{Te}}\sqrt{E_{De}/E}$, which can immediately produce the type III radio bursts and then be picked up by the main acceleration process, i.e., the seed electron population is formed by the runaway electrons, whose amount and, accordingly, the thermal-to-nonthermal energy partition, depends on the $E/E_{De}$ ratio. In case of super-Dreicer field, $E/E_{De}\gtrsim1$, most of available electrons will 'run away' and form the seed population, i.e., almost all these electrons are eventually accelerated in agreement with recent observations of the flare acceleration regions \citep{krucker_etal_2010, Fl_etal_2011}. Thus, we see that the helical component of the turbulence is easily capable of forming a seed population of electrons, which is needed for the mechanism of stochastic acceleration to efficiently work.

Consider now how this electric field can affect formation of the ion seed populations. For the estimate let us adopt a three-component plasma consisting of electrons ($e$), protons ($p$), and one more sort of ions ($i$) with the charge $Z|e|$, mass $m_i$, and number density $n_i$ so that $n_e=n_p+Zn_i$, permitted by a sub-Dreicer uniform electric field $\vE$. To address the question of the seed population formation we have to estimate the value and direction of the component flow velocities in the given plasma. To do so we have to determine conditions providing the balance between the electric force and forces of the dynamic friction between the plasma components driven by the Coulomb collisions between the particles.

The dynamic friction force acting from a plasma component $a$ on a given 'test' particle is calculated from the averaged momentum exchange between this plasma component and the test particle \citep{Trubnikov_1965}, i.e., 
\begin{equation}\label{Eq_fric_F_def}
    \vcF_a(\vu)=-\frac{ Z^2Q}{\mu_a}\int\frac{\vu-\vv}{|\vu-\vv|^3}f_a(\vv)d^{\,3}v,
\end{equation}
where $\vu$ is the velocity of the test particle, $\mu_a=Mm_a/(M+m_a)$ is the reduced mass defined by the test particle mass $M$ and $a$-component particle mass $m_a$, $f_a(\vv)$ is the distribution function of the component  $a$, and $Q=4{\pi}e^4\ln\Lambda_C$. To find the force acting on a 'mean' test particle we have yet to average force (\ref{Eq_fric_F_def}) over distribution function of the 'test' particle component. Both these averagings are convenient to perform assuming $\vu=\vu_T+\delta\vu$ and $\vv=\vv_T+\delta \vv$, where $\vu_T$ and $\vv_T$ are the corresponding thermal velocities with zero means (but nonzero rms values), while $\delta\vu$ and $\delta\vv$ are flow velocities driven by external electric field. Since we adopted $E/E_{De}<1$, the denominators are defined by the thermal velocities, whose contributions to the numerators are zeros, so the numerators are solely determined by the flow velocities.

Accordingly, the balance of forces acting on the electron component yields (we drop $\delta$ and use $\vv$ for all $\delta \vv$ below for short)
\begin{equation}\label{G_1b}
      -|e|\vE-\frac{Qn_p}{m_ev_{Te}^3}(\vv_e-\vv_p)-\frac{Z^2Qn_i}{m_ev_{Te}^3}(\vv_e-\vv_i)=0
\end{equation}
where we took into account that $v_{Te}{\gg}v_{Tp}$, $v_{Te}{\gg}v_{Ti}$  and $\mu_a{\approx}m_e$.
Then, in a similar way, we write for protons

\begin{figure}\centering
\includegraphics[width=0.8\columnwidth]{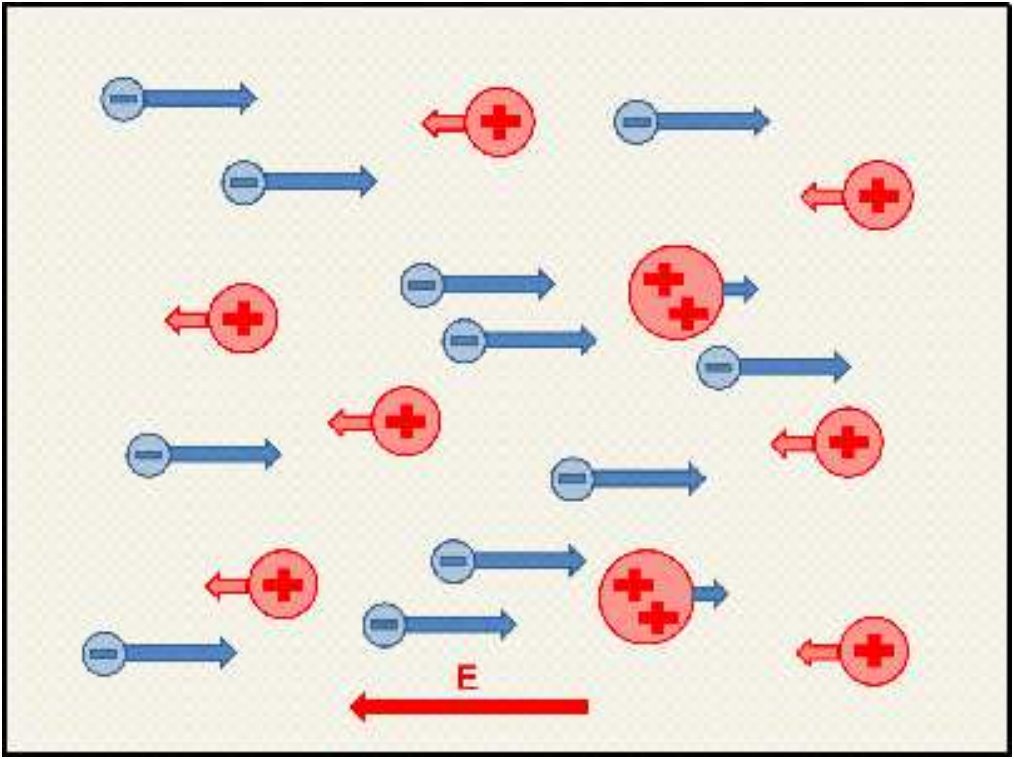}\\
\caption{\label{Fig_Dr_He} Illustration to the origin of the heavy ion velocity directed oppositely to the external electric field due to dynamic friction from moving electrons.  }
\end{figure}

\begin{equation}\label{G_1protons}
    |e|\vE-\frac{Qn_e}{m_ev_{Te}^3}(\vv_p-\vv_e)-\frac{Z^2Qn_i}{m_{ip}v_{Tp}^3}(\vv_p-\vv_i)=0,
\end{equation}
where we adopted for simplicity $v_{Tp}{\gg}v_{Ti}$; 
this approximation is the least accurate for the Helium ions but even is this case the error is within 30\%, which is acceptable for the purpose of the estimate; with the same accuracy we use below the proton mass $m_p$ for the reduced ion-proton masses $m_{ip}$.
Then, we add up Eqns (\ref{G_1b}) and (\ref{G_1protons}) to eliminate the electric field:
\begin{equation}\label{G_1sum}
      -\frac{QZ(Z-1)n_i}{m_ev_{Te}^3}\vv_e-\frac{Z^2Qn_i}{m_{ip}v_{Tp}^3}\vv_p+\frac{Z^2Qn_i}{m_{ip}v_{Tp}^3}\vv_i=0,
\end{equation}
where the plasma quasineutrality $n_e=n_p+Zn_i$ is taken into account and some small terms are discarded,
and then eliminate the electron velocity using the momentum conservation law $\vv_e=-(n_pm_p\vv_p+n_im_i\vv_i)/(n_em_e)$ (provided that the plasma is immobile as a whole in this reference frame)\footnote{We note that in a purely electron-proton plasma the momentum conservation would imply a very small proton flow velocity $\vv_p=-(m_e/m_p)\vv_e$.}:
\begin{equation}\label{G_1p4}
    \left[1+\frac{m_{ip}v_{Tp}^3}{m_ev_{Te}^3}\frac{Z-1}{Z}\frac{n_im_i}{m_en_e}\right]\vv_i=\left[1-\frac{m_{ip}v_{Tp}^3}{m_ev_{Te}^3}\frac{Z-1}{Z}\frac{n_pm_p}{m_en_e}\right]\vv_p.
\end{equation}
This equality demonstrates a remarkable property of the admixture ion behavior in the external electric field: the admixture ions move in the direction \textit{opposite}  to the main ions (since the second, negative term dominates the rhs for the natural abundance of the protons), i.e., the admixture ions move \textit{together} with electrons in the direction \textit{opposite} to the electric vector direction. This anomalous behavior noted earlier by \citet{Gurevich_1961, Furth_Rutherford_1972, Holman_1995} originates from the fact that the dynamic friction force produced by moving electron component on the ions with $Z>1$, and so proportional to $Z^2$, turns out to overcome the electric force, which is proportional to $Z$. Another remarkable property revealed by Eqn (\ref{G_1p4}) is that a significant ion flow velocity can only be achieved for a relatively tiny ion population for which the second term in the lhs brackets can be neglected compared with 1; for some abundant ions including helium and oxygen it is not the case since the second term is larger than or comparable to 1 even though $n_i/n_p\ll1$ for them.

Let us make estimates of the flow velocity for the Helium isotopes. Consider $^4He$ ($Z=2$ and $m_i=m_4=4m_p$) first and adopt $mv_T^2 \sim T$ for all plasma components, i.e.,  $v_{Tp}/v_{Te}\approx\sqrt{m_e/m_p}$, then
\begin{equation}\label{G_1p4_same_T}
    \left[1+2\frac{n_4}{n_e}\sqrt{\frac{m_{p}}{m_e}}\right]\vv_4=\left[1-\frac{n_p}{2n_e}\sqrt{\frac{m_{p}}{m_e}}\right]\vv_p.
\end{equation}

Since  $n_4/n_e\sim0.1\gg\sqrt{m_e/(4m_p)}\approx1/86\approx0.012$, for a rough estimate we can discard the first terms (ones) in both brackets, which simply yields   $\vv_4\sim-(n_p/4n_4)\vv_p$. To express the ion flow velocities via the electron flow velocity, however,  we have to calculate the ion velocity more accurately, i.e.,  retain small terms in Eqn (\ref{G_1p4_same_T}), which yields:

\begin{equation}\label{Eq_G_v4_vs_vp}
    \vv_4\approx-\frac{n_p}{4n_4}\left[1-\frac{n_e(4n_4+n_p)}{2n_pn_4}\sqrt{\frac{m_{e}}{m_p}}\right]\vv_p.
\end{equation}
Substituting Eqn (\ref{Eq_G_v4_vs_vp}) into the component momentum conservation law we find for the protons
\begin{equation}\label{Eq_G_vp_vs_ve}
 \vv_p\approx-\frac{2n_4}{n_p+4n_4}\sqrt{\frac{m_{e}}{m_p}} \vv_e,
\end{equation}
which is valid if ${2n_4}/({n_p+4n_4})>\sqrt{{m_{e}}/{m_p}}$ and, accordingly, it is noticeably larger than $\vv_p=-(m_e/m_p)\vv_e$ in the purely hydrogen plasma.
Now, substituting Eqn~(\ref{Eq_G_vp_vs_ve}) into Eqn~(\ref{Eq_G_v4_vs_vp}), we find for the $^4He$ ions

\begin{equation}\label{Eq_G_v4_vs_ve}
    \vv_4\approx   \frac{n_p}{2(n_p+4n_4)}\sqrt{\frac{m_{e}}{m_p}} \vv_e.
\end{equation}

A similar analysis for the other helium isotope, $^3He$ ($Z=2$ and $m_i=m_3=3m_p$) gives rise to essentially different result because of much lower $^3He$ number density, $n_3\approx1.7\cdot10^{-4}n_4$. For such a low density, we can safely discard the second term in the lhs of Eqn (\ref{G_1p4}) (while still discard 1 compared with the second term in the rhs), which eventually yields:

\begin{equation}\label{Eq_G_v3_vs_ve}
    \vv_3\approx\frac{n_pn_4}{n_e(n_p+4n_4)} \vv_e,
\end{equation}
roughly one order of magnitude larger than the $^4He$ flow velocity $\vv_4$, Eqn (\ref{Eq_G_v4_vs_ve}).

The results obtained allow estimating 'effective Dreicer' fields for various ions, which we define as an electric field needed to form the ion flow velocity equal to the thermal velocity of the same ion. Rewriting Eqns (\ref{Eq_G_v4_vs_ve}) and (\ref{Eq_G_v3_vs_ve}) in a compact form $\vv_i=K_i\vv_e$ and taking into account that $\vv_e=-v_{Te}\vE/E_{De}$ and $v_{Te}=v_{Ti}\sqrt{m_i/m_e}$ (we still assume equal temperatures of all plasma components), we obtain $\vv_i/v_{Ti}=-K_i\sqrt{m_i/m_e}\vE/E_{De}\equiv-\vE/E_{Di}$, where the latter equivalence is a definition of the effective ion Dreicer field. Then, using Eqn~(\ref{Eq_G_vp_vs_ve}) for protons and Eqns (\ref{Eq_G_v4_vs_ve}) and (\ref{Eq_G_v3_vs_ve}) for Helium ions we find

\begin{equation}\label{Eq_G_Dp_def}
    E_{Dp}\approx\frac{n_p+4n_4}{2 n_4} E_{De},
\end{equation}

\begin{equation}\label{Eq_G_D3_def}
    E_{D3}\approx\frac{n_e(n_p+4n_4)}{n_p n_4}\sqrt{\frac{m_{e}}{m_3}} E_{De},
\end{equation}

\begin{equation}\label{Eq_G_D4_def}
    E_{D4}\approx\frac{n_p+4n_4}{n_p} E_{De}.
\end{equation}
Adopting $n_4=0.1n_p$ (i.e., $n_e=1.2n_p$), we find $E_{D3}\approx0.23E_{De}$, while $E_{D4}\approx1.4E_{De}$ and $E_{Dp}\approx6E_{De}$, i.e., $E_{D3}{\ll}E_{De}<E_{D4}\ll E_{Dp}$.

This remarkable difference in the effective Dreicer fields implies that a given electric field can be sub-Dreicer for the electrons and  ions with large abundance (e.g., $p$ and $^4He$), while super-Dreicer for ions with small abundance (e.g., $^3He$), which is in particular the case for the electric field driven by the helical turbulence estimated by Eqn (\ref{Eq_Ereg_h_est}). Therefore, all available $^3He$ ions can 'run away' in the regular  electric field present in the helical turbulence and, thus, all the $^3He$ ions become available for further acceleration by this turbulence, while only a minor fraction of the $^4He$ ions from the corresponding Maxwellian tail will run away in the same electric field, so a much smaller number of the $^4He$ ions are available for acceleration, which offers an efficient mechanism of  $^3He$ enrichment in solar flares. Note, that given that $E_{D4}\ll E_{Dp}$, the same mechanism can be responsible for the $^4He/p$ enhancements discovered from the gamma-ray line study \citep{Share_Murphy_1998}.

\section{Discussion}

Currently, it is widely accepted that the stochastic mechanism plays a key role in charged particle acceleration in solar flares. Any stochastic acceleration model implies  a turbulent spectrum of waves (linear or nonlinear eigen-modes), whose resonant or nonresonant interactions with the charged particles result in the particle energy gain on average.
In this study we have found that the linear MHD modes excited on top of a \textit{twisted, nonpotential} mean magnetic field necessarily possess a nonzero \textit{kinetic helicity} and so differ from the standard MHD modes excited on top of the uniform mean magnetic field. Consequently, the turbulence composed of these 'modified' MHD modes will also possess a nonzero kinetic helicity.

We, therefore, have established that nonpotentiality of the magnetic field at the flare site (i.e., twists of the magnetic field lines) implies necessarily a certain level of helicity of the MHD turbulence created there by the primary flaring energy release. This turbulence helicity has potentially a number of highly important effects on the particle acceleration in the flares because, in addition to the common for the turbulence \textit{random} magnetic and electric fields, capable of particle scattering and stochastic acceleration respectively,  it also generates a nonzero mean \textit{regular} DC electric field with a rather significant value, see Eqn~(\ref{Eq_E_h}).

For a plasma obeying the standard  Ohm's law with Spitzer conductivity such an electric field would represent a problem since this would imply an unrealistically strong electric current, $J_c \sim 10^{18}$~A, and correspondingly strong magnetic field, $B_c\sim10^8$~G \citep[e.g.,][]{Emslie_Henoux_1995}. For the considered here helical turbulence the situation is entirely different. Indeed, the enhanced  turbulence reduces the plasma conductivity (by introducing an anomalous resistivity) and modifies the Ohm's law accordingly, see Appendix~II, and so the role of the enhanced, helicity-driven DC electric field is just to support the original electric current needed to maintain the original magnetic field twist ($\vj\propto \nabla\times\vB$), while not to enhance the current or generate additional magnetic field.

This electric current is formed by steadily moving thermal particles for which the effective electric conductivity derived in Appendix~II applies. However, because the electric field is enhanced, the amount of runaway electrons capable of leaving the thermal pool and transiting to a nonthermal population  increases accordingly. We emphasize that the helicity-driven DC electric field has almost no effect on particles with energy higher than 10s~keV, where standard stochastic acceleration remains dominant. {Accordingly, the particle spectra at $\gtrsim10$~keV will presumably have a standard power-law form as expected for the standard stochastic acceleration process.} In contrast, the DC field can affect and even dominate the particle acceleration at low energies---from the Maxwellian tail to a few or a few 10s~keV, where the rate of stochastic  particle acceleration     can be
greatly enhanced when the turbulence is helical, i.e. when
the helicity parameter $\alpha\neq0$; see estimates of Eqn~(\ref{10.49_estimate}).

There, depending on how strong the DC electric field is relative to the Dreicer field, it can control what fraction of the plasma particles will 'runaway', forming a seed particle population, and so go eventually into the accelerated component due to the standard stochastic acceleration of the seed particles. 
This process offers an efficient mechanism of the flaring energy partition between the thermal plasma and accelerated nonthermal particles {demanded by observations \citep[see, e.g.,][]{Altyntsev_etal_2012}}, which is in principle capable of providing any partition---from 'heating without acceleration' \citep{Battaglia_etal_2009} to 'acceleration without heating' \citep{Fl_etal_2011}.
These runaway electron populations  {can form  beams directed along the mean DC electric field, capable, in some cases, of producing} type III radio bursts and some of rapid fluctuations observed from the microwave and HXR radiations. {In fact, under conditions of  angular scattering a beam can only exist over roughly one mfp, since the beam-forming electrons will then isotropize. Therefore, if the beam produces or does not produce a type III burst  depends on how the mfp $\Lambda$ relates to the density scale, $L_d$: a drifting type III burst passing a broad range of frequencies can only be produced if $\Lambda \gg L_d$, while the beam will be destroyed before a significant spectral drift is achieved if $\Lambda \ll L_d$, thus, the spectral range of the corresponding radio emission must be rather narrow in the latter case, leading to a 'failed type III burst'. Perhaps, such failed type III bursts represent a widely observed phenomenon called 'metric spikes'\footnote{The metric spikes are narrowband spikes originating at 200--400~MHz at the frequency range adjacent to the starting frequency of the classical type III burst \citep[see, e.g., Figure~10 in][]{BBG}.} \citep{BBG, Benz_etal_1996}, which were proposed to come from the acceleration region directly. Indeed, if the level of turbulence is enhanced in the acceleration region and the mfp is reduced accordingly, the beams will be destroyed and the beam-driven instability responsible for the coherent radio emission quenched shortly after the beam formation; thus, the elementary radio burst will remain narrowband in contrast to formation of the broadband drifting structure in the classical type III bursts, which originate outside the acceleration region where the mfp is noticeably larger.     }

Furthermore, we point out that this same electric field can form highly non-even seed populations of various ions because of correspondingly distinct effective Dreicer fields of the ions with different abundances, which offers a natural way of understanding of the accelerated particle population enrichment by ions with relatively small abundances. This enrichment mechanism based on electrically selective extraction of the ions from their original thermal distributions with natural abundances, described by Eqn (\ref{G_1p4}), can account for the entire $^3He$ enhancement in the $^3He$-rich events, which have been puzzling for about 40 years by now {\citep[see, however,][]{Liu_etal_2006}}. Eqn (\ref{G_1p4}) shows that the ions with $Z>1$ and a low number density can have relatively small effective Dreicer fields, which is favorable for them to run away and be picked up by the bulk acceleration process.
In contrast, no enrichment is predicted for the $^2H$ ions, which are a comparably rare as the $^3He$ ions, but have $Z=1$ and so, for the hydrogen isotopes $^2H$ and $^3H$,  the terms containing $Z-1$ drop out from Eqn (\ref{G_1p4}) and these ions have exactly the same flow velocity as the protons {implying no enrichment for any of the hydrogen isotopes}. And indeed, no $^2H$ or $^3H$ enrichment has ever been observed from solar flares \citep{Koch_Koch_1984} {in agreement with the proposed here \textit{non-resonant enrichment mechanism}. Note, that although \textit{resonant enrichment mechanisms} \citep[see, e.g.,][]{Liu_etal_2006} are also capable of producing $^3He$ and other isotope enhancements, it is yet unclear how they are always so fine tuned to keep the $^2H$ ions unenriched in all cases.}

On the other hand, high-$Z$ ions can display even stronger enrichment than the $^3He$ ions, because $(Z-1)/Z\approx1$ for them in contrast to helium, where $(Z-1)/Z=1/2$. In fact, enrichment of $^3He$-rich events with ultra-heavy ions is widely observed \citep{Mason_2007}. Note also that a high correlation between the $^3He$-rich events and radio type III bursts \citep{Mason_2007} also receives a natural interpretation within the proposed enrichment mechanism, because the formation of the $^3He$ seed population happens along with formation of the runaway electrons responsible for  generation of the radio type III bursts. As has been said, whether the electron beam is formed or not is determined by the balance between the runaway electron acceleration by the electric field, which is proportional to $\Lambda\propto B_0^2/\langle {b}^2\rangle$  and their angular scattering by the same turbulence, which is reciprocal to $\Lambda$. Therefore, for a stronger turbulence, when, e.g., the induced electric field becomes closer to the $^4He$ Dreicer field, Eqn~(\ref{Eq_G_D4_def}), the electron angular scattering can effectively isotropize the electron beams and so quench the type III burst generation in more powerful flares not showing a considerable $^3He$ enrichment \citep{Mason_2007}, {although type III bursts can still be produced outside the acceleration region due to electron transport effects}.


Then, in the presence of this electric field the particle flux direction, according to Eqn~(\ref{10.45b}), depends now on their electric charge sign; thus, the accelerated electrons and protons are supposed to precipitate into the opposite foots of the same flaring loop or loop system, which offers a consistent explanation of the observed spatial displacement between sources of HXR (produced by electrons) and gamma-ray (produced by ions) emissions \citep{Lin_etal_2003, Hurford_etal_2006, Vilmer_etal_2011}. 


\section{Conclusions}

We have argued that the presence of strong
current helicity (twists of magnetic filed lines) implies a proportionally strong
helicity of the turbulence produced by release of the flaring energy
stored in the helical (nonpotential) magnetic field, which, in its turn, has highly important implications for the stochastic particle acceleration in solar flares. The proposed mechanism offers a nice consistent way of interpretation for (i) thermal to nonthermal energy partition in flares,  (ii) electron beam formation,  (iii) preferential acceleration of low-abundance isotopes with $Z>1$, and  (iv) observed spatial separation of the electron and proton precipitation sites.
Therefore, the stochastic acceleration by  helical turbulence is a promising
mechanism in the magnetized solar corona including the solar flares, jets, and eruptions.

It should be noted that this
attractive mechanism of particle  acceleration by helical turbulence
has not yet been analyzed in any detail, so many important aspects
of it remain unclear. Realistic self-consistent nonlinear models of
particle acceleration by the helical turbulence supplied by (yet
nonexistent) nonlinear theory of the helical turbulence are called
for to fully assess the role of this acceleration mechanism.

\section*{Acknowledgments}
This work was supported in part by NSF grants
AGS-0961867, AST-0908344, and NASA grants NNX10AF27G and NNX11AB49G to New Jersey
Institute of Technology,  RF Ministry of Education and Science grant 11.G34.31.0001,
and by the RFBR  grants 12-02-00173 and 12-02-00616.
The authors are greatly thankful to Valentina Abramenko and Sung-Hong Park for valuable discussions of the kinetic helicity on the Sun.

\section*{Appendix I\\ 
Helicity of MHD Eigen-Modes in a Plasma Permitted by a Nonpotential Magnetic Field}

Let us consider the kinetic helicity of the linear MHD modes and corresponding mean electric field $\vE_h=-\left<\vu\times\vb\right>/c$ in a plasma permitted by a nonpotential (and constant in time) force-free mean magnetic field based on the original MHD equations for the magnetic field and fluid velocity:
\begin{equation}\label{2.19}
\nabla{\cdot}\vB\,=\,0,$$$$ \frac{\pa\vB}{\pa
t}=\nabla\times[\vu\times\vB]+\nu_m\triangle\vB,
\qquad\nu_m=\frac{c^2}{4\pi\sigma}=\const ,\end{equation}
\begin{equation}\label{2.15}
  \rho\left(\frac{\pa\vu}{\pa
  t}+(\vu{\cdot}\nabla)\vu\right)\,=\,-\nabla  p + \vf
  +\frac{1}{4\pi}[\nabla\times\vB]\times\vB +
  \eta\triangle\vu + \frac{\eta}{3}\nabla(\nabla{\cdot}\vu),
  \end{equation}
where $\nu_m$ is the magnetic diffusivity, $\sigma$ is the electric conductivity, $\rho$ and $p$ are the plasma density and kinetic pressure,
$\vf$ is an external force (e.g., gravity), and $\eta$ is the dynamic viscosity. To investigate properties of weakly damping linear modes we neglect the dissipative terms entirely. Then, for simplicity we only consider the helicity originating from the magnetic field twist (the corresponding pseudoscalar is formed by the dot product of the polar vector $\vj$ and the axial vector $\vB$: $\vj\cdot\vB$, while no other  polar vectors are explicitly considered) and so we neglect the kinetic pressure and the external force assuming that their contribution to the helicity is smaller; so the equations read:
\begin{equation}\label{2.19a}
\nabla{\cdot}\vB\,=\,0,\qquad \frac{\pa\vB}{\pa
t}=\nabla\times[\vu\times\vB],\end{equation}
\begin{equation}\label{2.15a}
  \rho\left(\frac{\pa\vu}{\pa
  t}+(\vu{\cdot}\nabla)\vu\right)\,=\,  \frac{1}{4\pi}[\nabla\times\vB]\times\vB.
  \end{equation}

Now, to describe the small-amplitude linear modes satisfying these equations, we have to linearize them using Eqn~(\ref{10.39}) for the magnetic field and  adopting that $\vb$ and $\vu$ to be the first-order oscillating values of an MHD mode. Importantly, that upon substitution of Eqn~(\ref{10.39}) into Eqn~(\ref{2.15a}) we have to take into account that in the nonpotential force-free field $\nabla\times\vB_0=\alpha_{FFF}\vB_0\neq0$ unlike cases of a uniform field or nonuniform potential field, which yields 
\begin{equation}\label{1t}
    \frac{\partial\vb}{\partial t}=(\vB_0\cdot\nabla)\vu-(\vu\cdot\nabla)\vB_0,\qquad
    \frac{\partial\vu}{\partial t}=\frac{\alpha_{FFF}}{4\pi\rho}\vB_0\times\vb+\frac{1}{4\pi\rho}(\nabla\times\vb)\times\vB_0.
\end{equation}

Within the eikonal approximation (i.e., adopting the  wavelengths of the eigen-modes to be small compared with the source inhomogeneity scale)
 we can write $\vb=\vb_\omega e^{i\psi(\mathbf{r})-i\omega t}$ and a similar for $\vu$, which yields equations for the corresponding complex amplitudes:
\begin{equation}\label{2.19b}
\vb=(i/\omega)[i(\vB_0\cdot\nabla\psi)\vu-(\vu\cdot\nabla)\vB_0],
\end{equation}
\begin{equation}\label{2.15b}
\vu=\frac{i\alpha_{FFF}}{4\pi\rho\omega}\vB_0\times\vb-
\frac{1}{4\pi\rho\omega}[(\vb\cdot\vB_0)\nabla\psi-(\vB_0\cdot\nabla\psi)\vb].
\end{equation}

Let us solve these equations  to the  first order accuracy over the small parameter
$\alpha_{FFF}/|\nabla\psi|\ll 1$.
In the zeroth order approximation  we have
\begin{equation}\label{Eq_bu_eik}
\vb=-\frac{1}{\omega}(\vB_0\cdot\nabla\psi)\vu,\qquad\vu=-
\frac{1}{4\pi\rho\omega}(\vB_0\cdot\nabla\psi)\vb,
\end{equation}
which yields the eikonal
\begin{equation}\label{Eq_eik_0}
 \nabla_\parallel\psi=\pm\omega/v_A, \qquad v_A=B_0/\sqrt{4\pi\rho}.
\end{equation}
It is easy to see that in the zeroth approximation these perturbations are identical to the
purely alfv\'enic modes for which the conditions $\vb\cdot\nabla\psi=0,\
\vu\cdot\nabla\psi=0,\ \vb\cdot\vB_0=0$, and $\vu\times\vb=0$ are  satisfied.
Since we use the complex amplitudes the bilinear terms like
$ab$ must be computed as
$(1/2)\Re\langle ab^*\rangle$, where in addition to averaging over the period
$T=2\pi/\omega$ we also average over the  random phases of Fourier harmonics: 
\begin{equation}\label{2t}
\langle b_\mu b_\nu^*\rangle=(1/2)\langle b^2_\omega\rangle(\delta_{\mu\nu}-e^\parallel_\mu e^\parallel_\nu).
\end{equation}
In the zeroth over $\alpha_{FFF}$ approximation, the kinetic helicity parameter is zero: $\langle\vu\cdot(\nabla\times\vu^*) \rangle=0$.

Now, taking into account the first-order over $|\alpha_{FFF}/\nabla\psi|$ terms, Eqn~(\ref{2.15b}) yields
\begin{equation}\label{3t}
    \vu=-\frac{1}{4\pi\rho\omega}(\vB_0\cdot\nabla\psi)\vb+\frac{1}{4\pi\rho\omega}[i\alpha_{FFF}\vB_0\times\vb
    +(\vb\cdot\vB_0)\nabla\psi],
\end{equation}
where $\vb\cdot\vB_0\neq0$ and
\begin{equation}\label{4t}
   \nabla\times\vu=-\frac{i}{4\pi\rho\omega}(\vB_0\cdot\nabla\psi)(\nabla\psi\times\vb)+\frac{i}{4\pi\rho\omega}
   [-\nabla\psi\times(\vb\cdot\nabla)\vB_0+\vb\times(\vB_0\cdot\nabla)\nabla\psi+\vb\times(\nabla\psi\cdot\nabla)\vB_0].
\end{equation}
These two expressions allow calculating the kinetic helicity density
\begin{equation}\label{5t}
    \langle{h_k}\rangle=\frac{1}{2}\Re\langle\vu\cdot(\nabla\times\vu^*)\rangle=\frac{1}{4}\alpha_{FFF}v_A^2
    \frac{\langle b_\omega^2\rangle}{B_0^2},
\end{equation}
which is  nonzero any longer; it coincides with estimate~(\ref{Eq_Gyro_Turb_accel_h_k_FFF}) in the order of magnitude. 

\section*{Appendix II\\
Mean DC Electric Field Induced by the Helical Turbulence}

Let us now calculate the large-scale electric field, created by helical turbulence, taking into account the entire range of
harmonics forming the random fields of $\vu(\vr,t)$ and $\vb(\vr,t)$. From the first of Eqns~(\ref{1t}) we find
\begin{equation}\label{6t}
    \vb(\vr,t)=B_{0\nu}\int^t_{-\infty}\frac{\partial\vu(\vr,\tau)}{\partial x_\nu}d\tau-\frac{\partial\vB_0}{\partial x_\beta}
    \int^t_{-\infty}u_\beta(\vr,\tau)d\tau,
\end{equation}
which allows to form the required bilinear cross product $\vu\times\vb$
and perform the averaging
\begin{equation}\label{7t}
    \vE_h=-(1/c)\langle\vu(\vr,t)\times\vb(\vr,t)\rangle
\end{equation}
over the turbulent ensemble according to Eqn~(\ref{10.43}), which is convenient to express via the vector  components
\begin{equation}\label{8t}
    \langle u_\sigma(\vr,t)b_\mu(\vr,t)\rangle=-\frac{\partial B_{0\mu}}{\partial x_\beta}\int^t_{-\infty}
    \langle u_\sigma(\vr,t)u_\beta(\vr,\tau)\rangle d\tau + B_{0\nu}\int^t_{-\infty}\langle u_\sigma(\vr,t)
    \frac{\partial u_\mu(\vr,\tau)}{\partial x_\nu}\rangle d\tau.
\end{equation}
The first term in the rhs describes a modification to the Ohm's law due to the wave turbulence ensemble (anomalous resistivity), while the second one describes another correction to the Ohm's law due to the turbulence helicity; we will see below that these two terms provide comparable contribution in the problem studied.
Adopting for simplicity that the turbulence is statistically uniform and isotropic, we can express the first integral using the
invariant Kronecker's tensor as
\begin{equation}\label{9t}
    \int^t_{-\infty}\langle u_\sigma(\vr,t)u_\beta(\vr,\tau)\rangle d\tau=(\langle u^2\rangle\tau_c/3)\delta_{\sigma\beta},
\end{equation}
where $\langle u^2\rangle\tau_c/3=\nu_t$ is the magnetic turbulent diffusivity.

The second integral, which is a third rank tensor, is obviously proportional to the Levi-Civita's permutation
tensor $e_{\sigma\mu\nu}$ and  the kinetic helicity pseudoscalar $\alpha$
\begin{equation}\label{10t}
    \int^t_{-\infty}\left\langle u_\sigma(\vr,t)
    \frac{\partial u_\mu(\vr,\tau)}{\partial x_\nu}\right\rangle d\tau=\frac{1}{2}\alpha e_{\sigma\mu\nu}.
\end{equation}
Now,  the large-scale DC electric field can be written in the form
\begin{equation}\label{11t}
    \vE_h=(\nu_t/c)\nabla\times\vB_0-(\alpha/c)\vB_0=-(\alpha_{eff}/c)\vB_0,
\end{equation}
where, according to Eqns~(\ref{5t}) and (\ref{5.13})
\begin{equation}\label{12t}
 \alpha=-\frac{\tau_c}{12}\alpha_{FFF}v_A^2\frac{\langle b^2\rangle}{B_0^2},\qquad
    \alpha_{eff}=\alpha-(\nu_t/c)\alpha_{FFF}\end{equation}
and
\begin{equation}\label{13.t}
\alpha_{eff}=    -\frac{5\tau_c}{12}\alpha_{FFF}v_A^2\frac{\langle b^2\rangle}{B_0^2}
\end{equation}
is the total effective helicity parameter specifying the electric field in a nonpotential flux tube. Although both terms
in $\alpha_{eff}$ are comparable, the contribution from the anomalous magnetic diffusivity (anomalous resistivity) is four times larger than the direct helicity contribution. The correlation time can be estimated using the correlation scale $L_c$ and the Alfv\'en velocity $v_A$: $\tau_c=L_c/v_A$. Finally, recalling  that the magnetic field vector can be expressed via the electric current density, Eqn~(\ref{11t}) can be presented in the Ohm's law form $\vj=\sigma_{eff} \vE$, where $\sigma_{eff}= - (c^2/4\pi)(\alpha_{FFF}/\alpha_{eff})$ is the anomalous (turbulent) electric conductivity. When the effective conductivity is much smaller than the collisional Spitzer conductivity, having the same electric current (i.e., the same magnetic twist) requires a proportionally larger electric field, which is induced by the turbulence helicity.

\bibliographystyle{mn2e} 
\bibliography{fleishman,acceler,ms_bib} 

\begin{thebibliography}{}

\bibitem[\protect\citeauthoryear{{Abramenko}, {Wang} \&
  {Yurchishin}}{{Abramenko} et~al.}{1996}]{Abramenko_etal_1996}
{Abramenko} V.~I.,  {Wang} T.,    {Yurchishin} V.~B.,  1996, \solphys, 168, 75

\bibitem[\protect\citeauthoryear{{Abramenko}, {Wang} \&
  {Yurchishin}}{{Abramenko} et~al.}{1997}]{Abramenko_etal_1997}
{Abramenko} V.~I.,  {Wang} T.,    {Yurchishin} V.~B.,  1997, \solphys, 174, 291

\bibitem[\protect\citeauthoryear{{Altyntsev}, {Fleishman}, {Lesovoi} \&
  {Meshalkina}}{{Altyntsev} et~al.}{2012}]{Altyntsev_etal_2012}
{Altyntsev} A.~A.,  {Fleishman} G.~D.,  {Lesovoi} S.~V.,    {Meshalkina} N.~S.,
   2012, ArXiv e-prints

\bibitem[\protect\citeauthoryear{{Altyntsev}, {Fleishman}, {Huang} \&
  {Melnikov}}{{Altyntsev} et~al.}{2008}]{Altyntsev_etal_2008}
{Altyntsev} A.~T.,  {Fleishman} G.~D.,  {Huang} G.-L.,    {Melnikov} V.~F.,
  2008, \apj, 677, 1367

\bibitem[\protect\citeauthoryear{{Altyntsev}, {Grechnev}, {Meshalkina} \&
  {Yan}}{{Altyntsev} et~al.}{2007}]{Altyntsev_etal_2007}
{Altyntsev} A.~T.,  {Grechnev} V.~V.,  {Meshalkina} N.~S.,    {Yan} Y.,  2007,
  \solphys, 242, 111

\bibitem[\protect\citeauthoryear{{Aschwanden}}{{Aschwanden}}{2002}]{Aschw_2002}
{Aschwanden} M.~J.,  2002, {Particle Acceleration and Kinematics in Solar
  Flares}.
Particle Acceleration and Kinematics in Solar Flares, A Synthesis of Recent
  Observations and Theoretical Concepts, by Markus J.~Aschwanden, Lockheed
  Martin, Advanced technology Center, palo Alto, California, U.S.A.~Reprinted
  from SPACE SCIENCE REVIEWS, Volume 101, Nos.~1-2 Kluwer Academic Publishers,
  Dordrecht

\bibitem[\protect\citeauthoryear{{Aschwanden}}{{Aschwanden}}{2005}]{Aschwanden_2006}
{Aschwanden} M.~J.,  2005, {Physics of the Solar Corona. An Introduction with
  Problems and Solutions (2nd edition)}

\bibitem[\protect\citeauthoryear{{Aschwanden}, {Schwartz}, {Benz}, {Lin} \&
  {Pelling}}{{Aschwanden} et~al.}{1990}]{Aschwanden_etal_1990}
{Aschwanden} M.~J.,  {Schwartz} R.~A.,  {Benz} A.~O.,  {Lin} R.~P.,
  {Pelling} R.~M.,  1990, \solphys, 130, 39

\bibitem[\protect\citeauthoryear{{Aschwanden}, {Wills}, {Hudson}, {Kosugi} \&
  {Schwartz}}{{Aschwanden} et~al.}{1996}]{Aschwanden_etal_1996}
{Aschwanden} M.~J.,  {Wills} M.~J.,  {Hudson} H.~S.,  {Kosugi} T.,
  {Schwartz} R.~A.,  1996, \apj, 468, 398

\bibitem[\protect\citeauthoryear{{Bastian}, {Benz} \& {Gary}}{{Bastian}
  et~al.}{1998}]{BBG}
{Bastian} T.~S.,  {Benz} A.~O.,    {Gary} D.~E.,  1998, \araa, 36, 131

\bibitem[\protect\citeauthoryear{{Battaglia}, {Fletcher} \& {Benz}}{{Battaglia}
  et~al.}{2009}]{Battaglia_etal_2009}
{Battaglia} M.,  {Fletcher} L.,    {Benz} A.~O.,  2009, \aap, 498, 891

\bibitem[\protect\citeauthoryear{{Benz}, {Csillaghy} \& {Aschwanden}}{{Benz}
  et~al.}{1996}]{Benz_etal_1996}
{Benz} A.~O.,  {Csillaghy} A.,    {Aschwanden} M.~J.,  1996, \aap, 309, 291

\bibitem[\protect\citeauthoryear{{Browning} \& {Priest}}{{Browning} \&
  {Priest}}{1983}]{Browning_Priest_1983}
{Browning} P.~K.,  {Priest} E.~R.,  1983, \apj, 266, 848

\bibitem[\protect\citeauthoryear{{Bykov} \& {Fleishman}}{{Bykov} \&
  {Fleishman}}{2009}]{Byk_Fl_2009}
{Bykov} A.~M.,  {Fleishman} G.~D.,  2009, \apjl, 692, L45

\bibitem[\protect\citeauthoryear{{Emslie} \& {Henoux}}{{Emslie} \&
  {Henoux}}{1995}]{Emslie_Henoux_1995}
{Emslie} A.~G.,  {Henoux} J.-C.,  1995, \apj, 446, 371

\bibitem[\protect\citeauthoryear{{Emslie}, {Miller} \& {Brown}}{{Emslie}
  et~al.}{2004}]{Emslie_etal_2004}
{Emslie} A.~G.,  {Miller} J.~A.,    {Brown} J.~C.,  2004, \apjl, 602, L69

\bibitem[\protect\citeauthoryear{{Fedorov} \& {Stehlik}}{{Fedorov} \&
  {Stehlik}}{2010}]{Fedorov_Stehlik_2010}
{Fedorov} Y.,  {Stehlik} M.,  2010, Journal of Physics B: Atomic Molecular
  Physics, 43, 185701

\bibitem[\protect\citeauthoryear{{Fedorov}}{{Fedorov}}{2011}]{Fedorov_2011}
{Fedorov} Y.~I.,  2011, Kinematics and Physics of Celestial Bodies, 27, 1

\bibitem[\protect\citeauthoryear{{Fedorov}, {Kats}, {Kichatinov} \&
  {Stehlik}}{{Fedorov} et~al.}{1992}]{Fedorov_etal_1992}
{Fedorov} Y.~I.,  {Kats} M.~F.,  {Kichatinov} L.~L.,    {Stehlik} M.,  1992,
  \aap, 260, 499

\bibitem[\protect\citeauthoryear{{Fleishman}, {Kontar}, {Nita} \&
  {Gary}}{{Fleishman} et~al.}{2011}]{Fl_etal_2011}
{Fleishman} G.~D.,  {Kontar} E.~P.,  {Nita} G.~M.,    {Gary} D.~E.,  2011,
  \apjl, 731, L19

\bibitem[\protect\citeauthoryear{{Furth} \& {Rutherford}}{{Furth} \&
  {Rutherford}}{1972}]{Furth_Rutherford_1972}
{Furth} H.~P.,  {Rutherford} P.~H.,  1972, Physical Review Letters, 28, 545

\bibitem[\protect\citeauthoryear{{Grigis} \& {Benz}}{{Grigis} \&
  {Benz}}{2004}]{Grigis_Benz_2004}
{Grigis} P.~C.,  {Benz} A.~O.,  2004, \aap, 426, 1093

\bibitem[\protect\citeauthoryear{{Grigis} \& {Benz}}{{Grigis} \&
  {Benz}}{2006}]{Grigis_Benz_2006}
{Grigis} P.~C.,  {Benz} A.~O.,  2006, \aap, 458, 641

\bibitem[\protect\citeauthoryear{{Gurevich}}{{Gurevich}}{1961}]{Gurevich_1961}
{Gurevich} A.~V.,  1961, Sov. Phys. JETP, 13, 1282

\bibitem[\protect\citeauthoryear{{Hamilton} \& {Petrosian}}{{Hamilton} \&
  {Petrosian}}{1992}]{Petrosian92}
{Hamilton} R.~J.,  {Petrosian} V.,  1992, \apj, 398, 350

\bibitem[\protect\citeauthoryear{{Holman}}{{Holman}}{1985}]{Holman85}
{Holman} G.~D.,  1985, \apj, 293, 584

\bibitem[\protect\citeauthoryear{{Holman}}{{Holman}}{1995}]{Holman_1995}
{Holman} G.~D.,  1995, \apj, 452, 451

\bibitem[\protect\citeauthoryear{{Hurford}, {Krucker}, {Lin}, {Schwartz},
  {Share} \& {Smith}}{{Hurford} et~al.}{2006}]{Hurford_etal_2006}
{Hurford} G.~J.,  {Krucker} S.,  {Lin} R.~P.,  {Schwartz} R.~A.,  {Share}
  G.~H.,    {Smith} D.~M.,  2006, \apjl, 644, L93

\bibitem[\protect\citeauthoryear{{Jing}, {Park}, {Liu}, {Lee}, {Wiegelmann},
  {Xu}, {Deng} \& {Wang}}{{Jing} et~al.}{2012}]{Jing_etal_2012}
{Jing} J.,  {Park} S.~H.,  {Liu} C.,  {Lee} J.,  {Wiegelmann} T.,  {Xu} Y.,
  {Deng} N.,    {Wang} H.,  2012, \apjl, 752, L9

\bibitem[\protect\citeauthoryear{{Kichatinov}}{{Kichatinov}}{1983}]{Kichat_1983}
{Kichatinov} L.~L.,  1983, Sov. Phys. JETP Letters, 37, 51

\bibitem[\protect\citeauthoryear{{Kiplinger}, {Dennis}, {Frost}, {Orwig} \&
  {Emslie}}{{Kiplinger} et~al.}{1983}]{Kiplinger_etal_1983}
{Kiplinger} A.~L.,  {Dennis} B.~R.,  {Frost} K.~J.,  {Orwig} L.~E.,    {Emslie}
  A.~G.,  1983, \apjl, 265, L99

\bibitem[\protect\citeauthoryear{{Kocharov} \& {Kocharov}}{{Kocharov} \&
  {Kocharov}}{1984}]{Koch_Koch_1984}
{Kocharov} L.~G.,  {Kocharov} G.~E.,  1984, \ssr, 38, 89

\bibitem[\protect\citeauthoryear{{Kozlovsky}, {Murphy} \& {Share}}{{Kozlovsky}
  et~al.}{2004}]{Kozlovsky_etal_2004}
{Kozlovsky} B.,  {Murphy} R.~J.,    {Share} G.~H.,  2004, \apj, 604, 892

\bibitem[\protect\citeauthoryear{{Krause} \& {R\"adler}}{{Krause} \&
  {R\"adler}}{1980}]{KR80}
{Krause} F.,  {R\"adler} K.-H.,  1980, {Mean-field magnetohydrodynamics and
  dynamo theory (Oxford, Pergamon Press, Ltd., 271 p.)}

\bibitem[\protect\citeauthoryear{{Krucker}, {Hudson}, {Glesener}, {White},
  {Masuda}, {Wuelser} \& {Lin}}{{Krucker} et~al.}{2010}]{krucker_etal_2010}
{Krucker} S.,  {Hudson} H.~S.,  {Glesener} L.,  {White} S.~M.,  {Masuda} S.,
  {Wuelser} J.,    {Lin} R.~P.,  2010, \apj, 714, 1108

\bibitem[\protect\citeauthoryear{{Lin}, {Krucker}, {Hurford}, {Smith},
  {Hudson}, {Holman}, {Schwartz}, {Dennis}, {Share}, {Murphy}, {Emslie},
  {Johns-Krull} \& {Vilmer}}{{Lin} et~al.}{2003}]{Lin_etal_2003}
{Lin} R.~P.,  {Krucker} S.,  {Hurford} G.~J.,  {Smith} D.~M.,  {Hudson} H.~S.,
  {Holman} G.~D.,  {Schwartz} R.~A.,  {Dennis} B.~R.,  {Share} G.~H.,  {Murphy}
  R.~J.,  {Emslie} A.~G.,  {Johns-Krull} C.,    {Vilmer} N.,  2003, \apjl, 595,
  L69

\bibitem[\protect\citeauthoryear{{Liu}, {Petrosian} \& {Mason}}{{Liu}
  et~al.}{2006}]{Liu_etal_2006}
{Liu} S.,  {Petrosian} V.,    {Mason} G.~M.,  2006, \apj, 636, 462

\bibitem[\protect\citeauthoryear{{Longcope}, {Fisher} \& {Pevtsov}}{{Longcope}
  et~al.}{1998}]{Longcope_etal_1998}
{Longcope} D.~W.,  {Fisher} G.~H.,    {Pevtsov} A.~A.,  1998, \apj, 507, 417

\bibitem[\protect\citeauthoryear{{Mason}}{{Mason}}{2007}]{Mason_2007}
{Mason} G.~M.,  2007, \ssr, 130, 231

\bibitem[\protect\citeauthoryear{{Miller}}{{Miller}}{1998}]{Miller_1998}
{Miller} J.~A.,  1998, \ssr, 86, 79

\bibitem[\protect\citeauthoryear{{Miller}, {Cargill}, {Emslie}, {Holman},
  {Dennis}, {LaRosa}, {Winglee}, {Benka} \& {Tsuneta}}{{Miller}
  et~al.}{1997}]{Miller_etal_1997}
{Miller} J.~A.,  {Cargill} P.~J.,  {Emslie} A.~G.,  {Holman} G.~D.,  {Dennis}
  B.~R.,  {LaRosa} T.~N.,  {Winglee} R.~M.,  {Benka} S.~G.,    {Tsuneta} S.,
  1997, \jgr, 102, 14631

\bibitem[\protect\citeauthoryear{{Parker}}{{Parker}}{1979}]{Parker_1979}
{Parker} E.~N.,  1979, {Cosmical magnetic fields: Their origin and their
  activity (Oxford, Clarendon Press; New York, Oxford University Press, 858
  p.)}

\bibitem[\protect\citeauthoryear{{Petrosian}}{{Petrosian}}{2012}]{Petrosian_2012}
{Petrosian} V.,  2012, \ssr, p.~49

\bibitem[\protect\citeauthoryear{{Petrosian}, {McTiernan} \&
  {Marschhauser}}{{Petrosian} et~al.}{1994}]{Petrosian94}
{Petrosian} V.,  {McTiernan} J.~M.,    {Marschhauser} H.,  1994, \apj, 434, 747

\bibitem[\protect\citeauthoryear{{Reames}}{{Reames}}{1999}]{Reames_1999}
{Reames} D.~V.,  1999, \ssr, 90, 413

\bibitem[\protect\citeauthoryear{{Reid}, {Vilmer} \& {Kontar}}{{Reid}
  et~al.}{2011}]{Reid_etal_2011}
{Reid} H.~A.~S.,  {Vilmer} N.,    {Kontar} E.~P.,  2011, \aap, 529, A66

\bibitem[\protect\citeauthoryear{{Schlickeiser}}{{Schlickeiser}}{1989}]{Schlickeiser_1989}
{Schlickeiser} R.,  1989, \apj, 336, 243

\bibitem[\protect\citeauthoryear{{Share} \& {Murphy}}{{Share} \&
  {Murphy}}{1998}]{Share_Murphy_1998}
{Share} G.~H.,  {Murphy} R.~J.,  1998, \apj, 508, 876

\bibitem[\protect\citeauthoryear{{Toptygin}}{{Toptygin}}{1985}]{Toptygin_1985}
{Toptygin} I.~N.,  1985, {Cosmic rays in interplanetary magnetic fields
  (Dordrecht, D.~Reidel Publishing Co., 387 p.)}

\bibitem[\protect\citeauthoryear{{Trubnikov}}{{Trubnikov}}{1965}]{Trubnikov_1965}
{Trubnikov} B.~A.,  1965, Reviews of Plasma Physics, 1, 105

\bibitem[\protect\citeauthoryear{{Vainshtein} \& {Zel'dovich}}{{Vainshtein} \&
  {Zel'dovich}}{1972}]{Vainshtein_Zeldovich_1972}
{Vainshtein} S.~I.,  {Zel'dovich} Y.~B.,  1972, Soviet Physics Uspekhi, 15, 159

\bibitem[\protect\citeauthoryear{{Vajnshtejn}, {Zel'Dovich} \&
  {Ruzmajkin}}{{Vajnshtejn} et~al.}{1980}]{Vajnshtejn_etal_1980}
{Vajnshtejn} S.~I.,  {Zel'Dovich} Y.~B.,    {Ruzmajkin} A.~A.,  1980,
  {Turbulent dynamo in astrophysics (Moskva: Nauka, 352 p.)}

\bibitem[\protect\citeauthoryear{{Vilmer}}{{Vilmer}}{2012}]{Vilmer_2012}
{Vilmer} N.,  2012, Royal Society of London Philosophical Transactions Series
  A, 370, 3241

\bibitem[\protect\citeauthoryear{{Vilmer}, {MacKinnon} \& {Hurford}}{{Vilmer}
  et~al.}{2011}]{Vilmer_etal_2011}
{Vilmer} N.,  {MacKinnon} A.~L.,    {Hurford} G.~J.,  2011, \ssr, 159, 167

\bibitem[\protect\citeauthoryear{{Vlahos} \& {Raoult}}{{Vlahos} \&
  {Raoult}}{1995}]{Vlahos_Raoult_1995}
{Vlahos} L.,  {Raoult} A.,  1995, \aap, 296, 844

\bibitem[\protect\citeauthoryear{{Zharkova}, {Arzner}, {Benz}, {Browning},
  {Dauphin}, {Emslie}, {Fletcher}, {Kontar}, {Mann}, {Onofri}, {Petrosian},
  {Turkmani}, {Vilmer} \& {Vlahos}}{{Zharkova}
  et~al.}{2011}]{Zharkova_etal_2011}
{Zharkova} V.~V.,  {Arzner} K.,  {Benz} A.~O.,  {Browning} P.,  {Dauphin} C.,
  {Emslie} A.~G.,  {Fletcher} L.,  {Kontar} E.~P.,  {Mann} G.,  {Onofri} M.,
  {Petrosian} V.,  {Turkmani} R.,  {Vilmer} N.,    {Vlahos} L.,  2011, \ssr,
  159, 357

\end{thebibliography}
\label{lastpage}
\end{document}